\documentclass[12pt, a4paper]{article}

\usepackage[english]{babel}

\usepackage[a4paper,top=2.5cm,bottom=2.5cm,left=2.5cm,right=2.5cm,marginparwidth=1.75cm]{geometry}

\usepackage{amsmath}
\usepackage{graphicx}
\usepackage{comment}
\usepackage[colorlinks=false, allcolors=false]{hyperref}
\usepackage{float}
\usepackage{authblk}

\usepackage{setspace}

\renewenvironment{abstract}
  {\list{}{\rightmargin\leftmargin}%
  \item[\hspace{10mm}\textbf{\textsl{Abstract:}}]\relax}
{\endlist}

\graphicspath{{./figures/}}

\title{ECMAScript - The journey of a programming language from an idea to a standard}
\author{Juho Vepsäläinen}
\affil{Aalto University}
\affil{juho.vepsalainen@aalto.fi}
\date{}

\begin{document}
\maketitle

\begin{abstract}
\textsl{A significant portion of the web is powered by ECMAScript. As a web technology, it is ubiquitous and available on most platforms natively or through a web browser. ECMAScript is the dominant language of the web, but at the same time, it was not designed as such. The story of ECMAScript is a story of the impact of standardization on the popularity of technology. Simultaneously, the story shows how external pressures can shape a programming language and how politics can mar the evolution of a standard. In this article, we will go through the movements that led to the dominant position of ECMAScript, evaluate the factors leading to it, and consider its evolution using the Futures Triangle framework and the theory of standards wars.}
\end{abstract}


\textbf{Disclaimer}

This preprint has been released with the permission of \href{https://www.euras.org/conference/EURAS-2023-Conference}{EURAS 2023}. It includes minor stylistic tweaks compared to the original due to the technical constraints of arXiv.

\section{Introduction} \label{sec:introduction}

Based on \cite{w3techsUsageStatisticsFEJS}, JavaScript is used by 98.3\% of websites on the client side. \cite{w3techsUsageStatisticsBEJS} estimates the market share on the server side to be around 2\%. Especially the client side figure is considerable, and it tells about the success of JavaScript in web development. JavaScript is a story of success in standardization and its standardized form; ECMAScript\footnote {Naming makes a difference regarding trademarks. Due to this, the standard version is called ECMAScript and not JavaScript \cite{wirfs2020javascript}. In this article, we use the terms interchangeably.}, as the language is the backbone of web development and the most popular programming language of the world with the largest deployed base \cite{wirfs2020javascript}.

The benefits and drawbacks of standardization are well understood, as shown by \cite{blind2009standardisation} and \cite{swann2010economics, swann2010international}, for example. In the case of JavaScript, the benefits outweigh the drawbacks, and the programming language has become an instrumental part of the web economy. It can be argued this would not have happened without the successful standardization efforts.

The topic of standardization in programming languages has been covered for early programming languages such as FORTRAN, COBOL, and PL/1 by \cite{baird1977programming}, \cite{hudak1989conception} covers the topic for functional languages, \cite{stroustrup2007evolving, stroustrup2020thriving} for C++, and \cite{urma2017programming} on a more general level. \cite{wirfsprogramming2016} looks at the topic of programming language standardization from a practical point of view and explains well how standards bodies operate.


\cite{wirfs2020javascript} have covered the first twenty years of JavaScript in detail from its inception in 1995. The work gives a good starting point for a more focused study on understanding how standardization affected the development and adoption of JavaScript. Therefore, to scope the study, we have formalized the research question of the paper as follows:

\begin{verbatim}
What was the impact of standardization in the evolution of JavaScript?
\end{verbatim}

A part of the article has been summarized from \cite{wirfs2020javascript}, and we recommend reading it to understand the process and especially the technical side in greater detail. The contribution of this article is a qualitative evaluation performed using the \cite{inayatullah2008six}'s Futures Triangle framework and \cite{shapiro1999information}'s classification of standards wars for considering which factors were affecting a given era of JavaScript. The article opens the way for research within the web programming language space, especially for comparative studies.



To understand how the Futures Triangle and the standards wars classification work, we cover the ideas first in Section \ref{sec:methodology}. We discuss ECMAScript and its eras in detail in the following sections before discussing the findings in Section \ref{sec:discussion} and concluding the article in Section \ref{sec:conclusion}.

\section{Methodology} \label{sec:methodology}

\begin{figure}[H]
    \centering
    \includegraphics[scale=0.4]{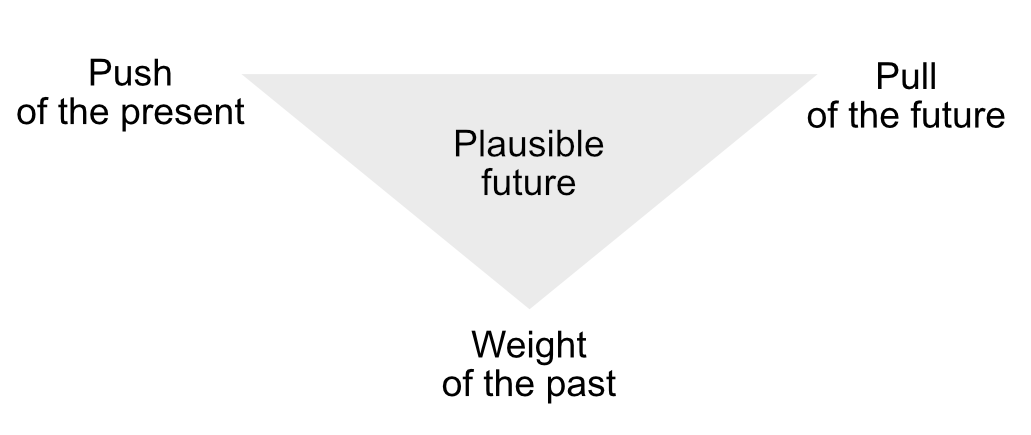}
    \caption{Plausible future \cite{inayatullah2008six}}
    \label{image:plausiblefuture}
\end{figure}

\cite{inayatullah2008six}'s Futures Triangle evaluates plausible futures through three aspects: the push of the present, the weight of the past, and the pull of the future. The push of the present contains current trends, known policies, and recent developments. Weight of the past represents obstacles that hold development back; for example, regulation or existing investments can be such. The pull of the future looks forward and helps to imagine possible futures. Brainstorming on these three factors allows discussion about what is possible and what is not, and why \cite{fergnani2020futures}. The framework is also useful for analyzing the past, although even then, it is speculative and up to an interpretation of the given situation. Figure \ref{image:plausiblefuture} shows how the three factors form a triangle in combination.

\cite{fergnani2020futures}'s Futures Triangle 2.0 expands the model by assigning a weight per axis and combining it with scenario planning. Using the expanded model, it is possible to evaluate scenarios relative to each other. In this case, the expansion does not provide much value, but it would be helpful if you project multiple plausible futures and evaluate them in detail.




\cite{shapiro1999information}'s typing of standards wars gives another axis for evaluation, allowing us to consider JavaScript relative to its contemporary technologies in a given era. The typing approaches the topic through compatibility with a rival technology and allows for considering compatibility with the approaches. Table \ref{table:shapiro} shows the model in detail.

\begin{table}[ht]
 \caption{Types of standards wars as described by \cite{shapiro1999information}}
 \begin{tabular}{ |p{4.9cm}|p{4.9cm} p{4.9cm}| }
 \hline
    & \multicolumn{2}{c|}{Rival technology}\\
 \hline
Your Technology & Compatible & Incompatible \\
Compatible & Rival evolutions & Evolution versus revolution \\
Incompatible & Revolution versus evolution & Rival revolutions \\
 \hline
 \end{tabular}
 \label{table:shapiro}
\end{table}

\section{Creation of JavaScript} \label{sec:creation}

The story of JavaScript is a story of a language that became the dominant one in web development, although it was never designed as one. It was flexible and good enough for the purpose, but you can argue it still inherits some of its original flaws. Over time, the understanding of these flaws has improved, however, and the language has been refined to address them to an extent. The surrounding conditions, i.e., the need for backward compatibility, have given constrained these refinements, but regardless JavaScript we write today is a somewhat different language than what it was in the beginning.

\subsection{Background}

Since the introduction of Mosaic, one of the first web browsers, in 1993, the web has become portable across different desktop environments (Windows, Macintosh, Unix). The wide support gave a potential platform for the development of cross-platform applications \cite{severance2012javascript}. As \cite{leprohon2017ecmascript} points out, web designers and developers did not have experience with Java or object-oriented principles, yet a small scale \textbf{scripting language} which could run in the browser was needed. In 1995, Netscape Communications decided to introduce this type of language for the web. Through the influence of Sun Microsystems and its popular Java language, it was mandated that the language should look like Java, ruling out certain designs, such as BASIC, while encouraging Java-like design choices \cite{wirfs2020javascript}.

The task of creating the first version of JavaScript was given to Brendan Eich, who had prior experience easy to use and pedagogical languages, such as HyperTalk, Logo, and Self \cite{severance2012javascript}. The influence of these languages can be seen in the design of JavaScript, and it is likely because it involves several, perhaps, nonconventional design choices. Eich famously completed the first language version in ten days in May 1995 \cite{wirfs2020javascript}.

The initial version was exposed to the public using the name LiveScript in September 1995 as a part of a beta version of Netscape Navigator 2.0. It was renamed as JavaScript in the stable version in March 1996 \cite{wirfs2020javascript}. The naming choice was a source of confusion as JavaScript did not have much to do with Java as a language \cite{leprohon2017ecmascript}. The first version of JavaScript was incomplete relative to the complete language design and included several bugs Brendan Eich continued to fix during subsequent Netscape Navigator 2.0 releases \cite{wirfs2020javascript}.

\subsection{Vision for JavaScript}

The initial vision was that JavaScript would complement Java so that functionality written in JavaScript could be combined together with Java applets and other components \cite{wirfs2020javascript}. Perhaps ironically, JavaScript does not have much to do with Java in terms of language design. However, JavaScript inherits several design decisions at the API level from Java 1.0, some of which, such as the Date module, are questionable today. JavaScript 1.0 was modeled after the syntax of C with inspiration from AWK. Still, in contrast to C, JavaScript statements can occur outside of a function's body, making it more flexible, not to mention many other subtle differences and, most importantly, dynamic vs. static typing of C \cite{wirfs2020javascript}.

It is easy to argue that the initial design of JavaScript was not perfect, as even admitted by the creator in \cite{eich2008}. Many of the early issues still plague the language even today, as in programming language design, it is easy to add features but difficult or even impossible to remove them without a significant revision. For a language with a broad installed base, publishing a breaking version could lead to unintended breakage, and we will discuss a concrete case of this later in the article as we cover so-called smooshgate. The issue also came up with the never released ES4, which we will touch on in a later section.

\subsection{JScript - a competitor appears}


To maintain feature parity with Netscape Navigator, Microsoft developed an interpreter called JScript for the language \cite{wirfs2020javascript}. Microsoft also implemented Visual Basic Script (VBScript) to build on its technology. Having to support JavaScript was an unhappy but necessary compromise to gain market share.

Given there was no official specification for JavaScript, Microsoft was forced to reverse engineer its implementation based on what Netscape did to remain competitive \cite{wirfs2020javascript}. Until a standard could emerge, the specification lived in Netscape's implementation of the language, which was not open source and available to study at the time.

\subsection{Discussion}

Netscape created JavaScript language to meet the demands of the interactive web while giving a competitive advantage to Sun Microsystems and Netscape. In the Futures Triangle, this represents \textbf{push of the present} and was driven by a \textbf{pull of the future} of a vision where client-side scripting is combined with server-side code, namely Java. Due to the constraints, the project already had some \textbf{weight of the past}, and this is visible in naming and high-level decisions that affected the language design.

Out of these three factors, it seems like the push of the present was the main driver for development as the web was still in its early phases, and it was a good time to influence its direction. At the same time, the development was motivated by competitive pressures as Netscape did not exist in a vacuum, as we will learn in the following section.

JavaScript had early competition in the form of JScript. In \cite{shapiro1999information}'s model, the situation represents a case of \textbf{rival evolutions} as JScript had to maintain compatibility with JavaScript for Internet Explorer to be able to capture market share.

The creation of JavaScript led to the birth of a new market. Had Brendan Eich known what was to come, he would likely have convinced Netscape to put more effort into the initial design. At the same time, the initial design was good enough to become popular beyond what could have been anticipated. The popularity of JavaScript followed the popularity of the web and was one of the key drivers of its growth as it became the fuel for developing web applications.



\section{From JavaScript to the first ECMAScript standard}

As Netscape shepherded the initial development of JavaScript and Microsoft followed with their own version of it in the form of JScript, there was clear pressure to standardize. Given Netscape leadership was sensitive to criticism from Microsoft, they wanted Brendan Eich to work on a specification in 1996, and that is what he did as he rewrote the initial JavaScript engine Mocha as a new one called SpiderMonkey and extracted a specification for JavaScript 1.2 \cite{wirfs2020javascript}.



From the beginning of the JavaScript project in 1995, it was clear that the web needed an open scripting standard, and the pressure from Microsoft was a sign of this. The target of Netscape and Sun Microsystems was to find a venue to collaborate with Microsoft while avoiding domination through their technology offerings, such as VBScript. W3C and IETF were potential early candidates for a standards body, but neither was the right fit for JavaScript. Through a personal connection, Netscape reached out to Ecma, and a fit was found. \cite{wirfs2020javascript}




\subsection{The first meeting of TC39}

The first meeting of technical committee 39 (TC39) organized in September 1996 was attended by thirty members, and David Stryker from Netscape proposed that the target should be to create a specification with minimal deviations from the then-current implementations and any language extensions should be deferred to the future to capture the current state of the language. \cite{wirfs2020javascript}

Thomas Reardon from Microsoft recommended the committee avoid duplication by leaving the standardization of an HTML object model to W3C because Netscape's and Microsoft's core features were similar, but the HTML APIs were different. These early constraints shaped the work of TC39 to develop only platform-independent features, which remains the group's core principle to this day. The point was to have a formal language definition while leaving space for competition outside of it. The approach has worked well in practice, and competition has occurred, for example, in interpreter performance. \cite{wirfs2020javascript}

\subsection{Creating the first draft for a standard}

\cite{wirfs2020javascript} describes the early struggles related to creating the initial specification in great detail, and it is a story of politics and intrigue. Essentially different parties, namely Netscape, Borland, and Microsoft, had their own specifications, which had to be remedied into a single, shared one as the first standard. The question was which specification to use for further refinement. Since Ecma was used to working with Microsoft Word, they started working based on Microsoft's version of the initial specification instead of Netscape's.

The target of the specification work was set to January 1997 for the initial draft and April 1997 for the final one. Any feature in all three proposals was considered uncontroversial during the work, while other points needed reconciliation. Any unique features were moved to Proposed Extensions to consider later. Another essential point was to avoid changes forcing altering existing applications, which became a guideline for future standard editions. \cite{wirfs2020javascript}

\subsection{Completion of the first draft}

The first draft of JavaScript from the 10th of January 1997 defined the language, and many of the definitions exist in the specification even today \cite{wirfs2020javascript}. One of the unique features of the definition was pseudocode-style definition language features to allow understanding of how the language works. Pseudocode works as a medium between written language and programming, providing an excellent way to understand the content without learning a specific language.

To finalize the standard before the April deadline, TC39 met regularly. The group wrote a test case for any unclear edge case and used it against the current runtimes to capture behavior. If behavior differed between implementations, consensus had to be found, and the behavior had to be specified. Some decisions have implications even today, leading to specific programming idioms visible in JavaScript programs. The group missed its April deadline by a month and completed the work 2nd of May 1997. The resulting document became ECMA Standard 262, or \href{https://www.ecma-international.org/publications-and-standards/standards/ecma-262/}{ECMA-262} for short. After minor editing, it was submitted to the ISO fast-track process, and the first version of the standard (ES1) was published on the 10th of September 1997. \cite{wirfs2020javascript}

\subsection{Discussion}

JavaScript was standardized as ECMAScript to address the pressures faced by the growing web ecosystem. At the same time, standardization was a way to even the playing field and gave precedence to the open web. The alternative was a web driven by proprietary platforms and technologies, and standardization of JavaScript did not remove this option entirely. The need for a standard is a clear example of a \textbf{push of the present} as it had to capture the behavior of the existing implementations that represented the \textbf{weight of the past}. The outcome of the standardization effort was motivated by \textbf{pull of the future} of an early vision for an interoperable language for the web.

Out of these factors, the push of the present stands out as something important, as there was a clear need to standardize. Pull of the future had some impact, but it feels like the acute situation had more weight. With ES4, the lack of a clear vision became a true problem, as we will see later.

The creation of the ECMAScript standard signifies an end to early JavaScript-related standards wars. It allowed browser vendors to align behind a single client-side scripting standard.


\section{From ES1 to ES3}

After the completion of ES1, both Netscape and Microsoft continued work on their own and added language extensions to their implementations. Meanwhile, the official TC39 meetings had changed form into management and strategy sessions, while most of the technical work was done in informal technical working groups. In July 1997, TC39 agreed on the next steps toward Version 2, and the official target for its completion was set to December 1997. In September 1997, it was agreed that the new version should retain backward compatibility with its earlier version. \cite{wirfs2020javascript}

\subsection{ES3 - naming and mode of working}

Given TC39 and ISO moved at different speeds, two groups with overlapping memberships worked on two separate specification drafts for Edition 2 and Version 2. To solve the dilemma, it was decided that the next version of ECMAScript should jump straight to 3. For this reason, ES2 never existed. \cite{wirfs2020javascript}

Many of the language extensions developed by Netscape and Microsoft made their way to the ES3 standard, and some were dropped or adopted in a later edition. As with ES1, the target of ES3 was to continue specifying already implemented features while adding a few not existing anywhere yet requiring design work not needed before. The work proceeded in monthly face-to-face meetings to resolve proposals and open issues to solidify the standard. For more complex tasks, such as internationalization (i18n), TC39 set up specific working groups. \cite{wirfs2020javascript}

Compared to ES1, the work proceeded slower, and by Spring 1999, it was clear the work could not be completed by June, but December was possible. In March, decisions were made to cut or defer features to catch the deadline. The initial draft was completed in August, and after identifying and fixing minor issues, the final draft was handed over on October 1999. After final changes, including a significant backward compatibility fix, ES3 was approved on December 1999 by ECMA, and it took until 2009 for the next version. \cite{wirfs2020javascript}

\subsection{Adoption of ES3}

As pointed out by \cite{wirfs2020javascript}, at that time, updating web browsers was not as common as now, and as a result, it was only around 2009 that web developers could expect ES3 features to be broadly supported. Later on, this dilemma was solved by tooling and browser vendors to a large extent. Compared to 1999 or even 2009, the current pace of development is significantly faster, and as we will see in the following sections, now the specification receives yearly updates. Before we could get there, the situation was the opposite for a long while.

\subsection{Discussion}

The shift from ES1 to ES3 was motivated by \textbf{push of present} in need to integrate browser-specific language extensions to the standard. During the standardization process, there was also awareness of new features to specify and add. At the same time, there was \textbf{weight of the past} in the form of backward compatibility to maintain. A shared vision of scripting on the web represented the \textbf{pull of the future}.

Out of the factors, the push of the present was the dominant one, as the evolution of the web platform put pressure on the standard to improve and catch the changes while allowing increased compatibility between the browsers. Compatibility was an essential factor for browser vendors as it allowed them to remain competitive while serving the growing web user base.




\section{ES4 - the version that never was}

According to \cite{wirfs2020javascript}, the perception of JavaScript changed over time as a language that had been initially perceived as a sidekick of Java had taken a direction of its own, and web developers began to understand it would be the only language they would need. Through increased usage, educators and evangelists appeared around JavaScript and they began to change its public perception. In addition to promotion, these people developed initial tooling to support JavaScript development in larger-scale use cases and data interchange format JSON that later on became an important part of web development overall.


With the introduction of DHTML and XMLHTTP by Microsoft in 1997, it became possible to develop interactive and dynamic web applications without having to refresh the page \cite{wirfs2020javascript}. Although the idea was used in limited context after that, it remained hidden from the public until 2005, when the technique was coined as AJAX that enabled the rise of JavaScript-centric Single Page Applications (SPAs) a few years later \cite{severance2012javascript,woychowsky2006}. Technical development displaced the earlier vision where Java applets and JavaScript were used together and represented a major technological landscape shift.

\subsection{ES4 - imagining the future of web development}

The background gives a good starting point for understanding why the ES4 standardization effort failed to materialize. By the early 2000s, it was clear the web was a winning technology, and it was time to shape its future. Compared to the earlier standardization rounds, ES4 had to imagine the future as the browser vendors could not initially provide features to standardize anymore. There was also confusion within TC39 about whether a breaking version fixing past mistakes was needed or whether iterative enhancement would be the way to go. Brendan Eich, the creator of the language, advocated for a breaking version, while a fragment of the group preferred an iterative approach (ES3.1). To make clear the new standard would be breaking, it was called ECMAScript 2.0 (ES4). \cite{wirfs2020javascript}

The effort towards ES4 started in early 1998, and it was filled with proposals improving the language on several fronts \cite{wirfs2020javascript}. Interestingly enough, some of these features, such as optional type annotations, are still under discussion and may eventually make their way to the language.


\subsection{Microsoft's success in the browser wars and its impact on standardization}

\begin{figure}[H]
    \centering
    \includegraphics[scale=0.12]{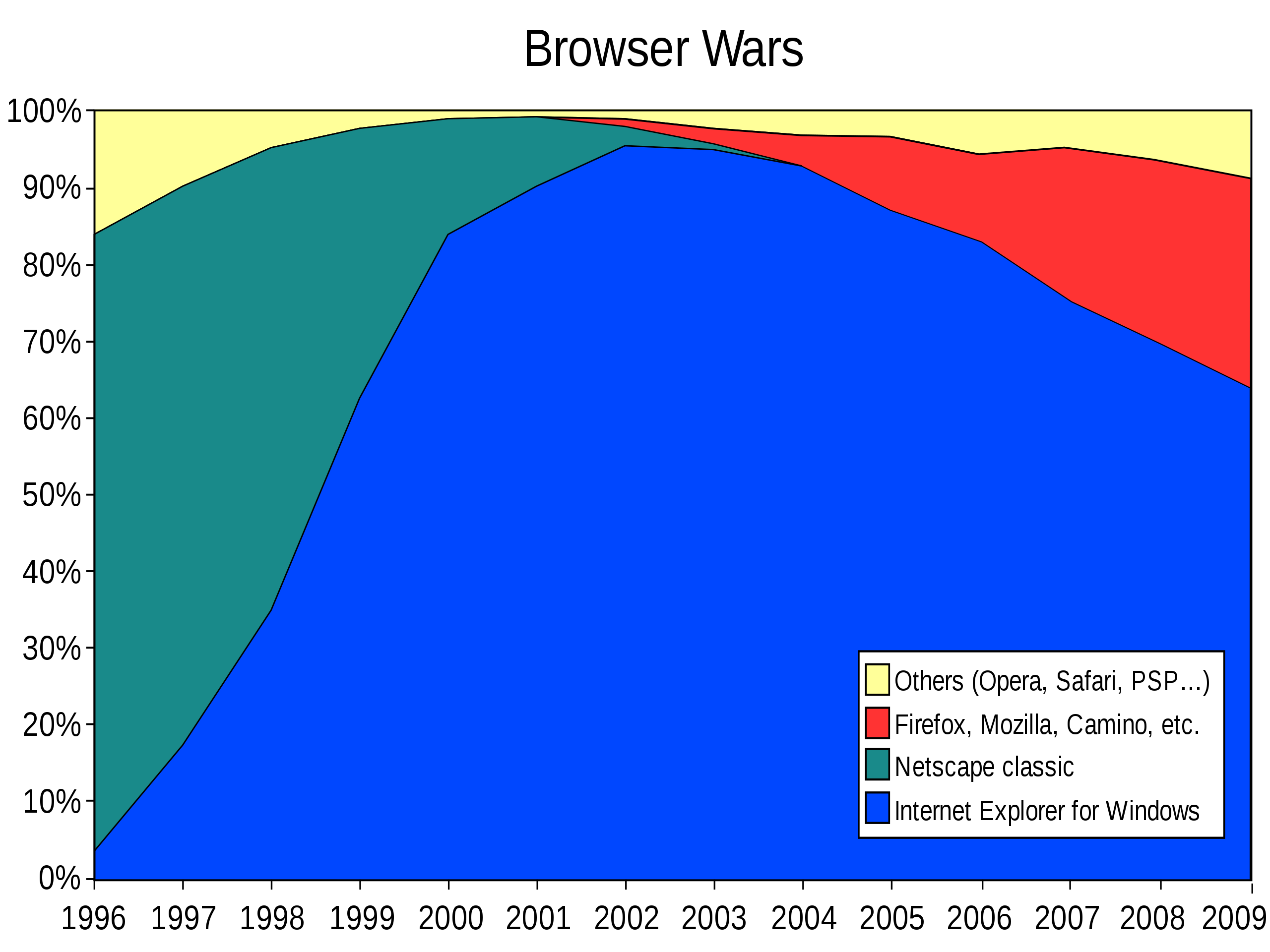}
    \label{fig:browser-wars}
    \caption{\href{https://commons.wikimedia.org/w/index.php?curid=1128061}{By Wereon - Own work based on: Browser Wars.png}, CC BY-SA 3.0}
\end{figure}

According to \cite{wirfs2020javascript}, the announcement of JScript.NET was the beginning of the end for the ES4 effort as simultaneously Microsoft won the browser wars and attained a market share of over 90\% for its Internet Explorer browser as seen in Figure \ref{fig:browser-wars}. The shift in the market meant that Microsoft could focus instead of the open web towards its own technologies to replace the incumbent solutions with its own proprietary alternatives. 

\subsection{Impact of ActionScript}

During the same era, there were several other attempts at improving the language. The most notable one is ActionScript from Macromedia's Flash authoring tool, which was a dialect of ES3 and did not follow the specification strictly. Flash became particularly popular in the early 2000s and was used to implement heavily interactive user experiences. Eventually, Macromedia developed a virtual machine and ran a simplified version of the proposed ES4 against it as ActionScript 3.0. Before the completion of the effort, Macromedia had been bought by Adobe, and the end result was known as Adobe Flash. \cite{wirfs2020javascript}

To help move ES4 forward, Macromedia (now owned by Adobe) joined Ecma in November 2003 as this helped to align the design of ActionScript with the future specification. In early 2004, it was clear to Brendan Eich, then Mozilla's CTO, that the future of the open web was threatened by proprietary platforms such as .NET, Flash, and others. \cite{wirfs2020javascript}


\subsection{ES3.1 vs. ES4}



As pointed out by \cite{crockford2012javascript}, when he joined TC39, it did not contain members using the language, as the only members were browser vendors. Crockford was not convinced that a new version of the language would solve any problems but found Allen Wirfs-Brock as an ally from Microsoft and was able to convince him that an iterative approach to language design would be the preferable way to go \cite{wirfs2020javascript}. The collaboration resulted in an alternate proposal that came to be known as ES3.1, and it proceeded parallel to the ES4 effort \cite{wirfs2020javascript}.

As the ES4 effort kept going and timelines shifted, Adobe, one of the key drivers, decided to step out in 2008. Adobe stepping out meant the ES4 effort was over and marked the beginning of the harmonization of ECMAScript. After harmonization, the plan was to focus on completing ES3.1 while simultaneously collaborating on a follow-up called Harmony which would not be constrained by decisions made for ES4. \cite{wirfs2020javascript}


\subsection{Discussion}

The failure of ES4 to materialize as a standard provides a good example of how standardization can fail. During ES4, TC39 worked to design a new language that would fix past mistakes rather than standardize existing behavior like before. As noted by Crockford, the failure of ES4 had to do with the fact that standards bodies should specify, not design \cite{wirfs2020javascript}. Design is a task suitable for the community, and as we will see later, TC39 has adopted a way of working that enables this, meaning TC39 can focus on specifying to avoid the fate of ES4.

Compared to the earlier standardization rounds where existing behavior was specified as a standard, the ES4 effort tried to re-imagine the whole language and fix its past mistakes (\textbf{push of the present}). At the same time, there was a need to support past features and existing browsers as a form of \textbf{weight of the past}. The overall vision of JavaScript as a programming language for the web represented the \textbf{pull of the future}.

Although there was a clear need to improve the language, a revolutionary approach failed, and instead, the standardization efforts took an evolutionary path as seen in the following sections. It is good to remember the context in which JavaScript exists, especially on the web. To develop applications for the web platform, you have to consider the variety of browsers available to the users. For JavaScript to improve on the web platform, the language evolution has to consider these environmental factors. Due to the constraints of the web and its varied audience, maintaining backward compatibility is important, and for this reason, the evolutionary approach was more natural for JavaScript rather than going for a complete language revolution that would risk browser support. From a modern point of view, JavaScript has become a compilation target for other languages, and we will discuss this point later in detail as we touch on TypeScript and the topic of transpilation.

With ES4, standards wars had begun again as JavaScript alternatives had begun to appear on the web. Adobe's Flash is a good example of a strong contender, and it represented a rival technology that was not compatible with JavaScript, therefore, matching the case of \textbf{rival revolutions} in \cite{shapiro1999information}'s model.

The reason why it was difficult for Flash and other alternatives to compete with JavaScript in the long term has to do with the way JavaScript code is delivered to the clients. As the browser contains the interpreter already, no additional runtime is required. That was not the case with the alternatives and having this step added to the friction of using the competing technologies. As web technologies evolved in the open, the situation became worse for the options, and solutions like Flash were phased out as the platform caught up with its capabilities.


\section{ES5}

Given the ES4 standardization effort failed, ECMAScript changed its direction with a renewed focus that aimed to finish ES3.1 while planning a more significant follow-up codenamed Harmony. After done, ES3.1 came to be known as ES5 and eventually, many of the features planned for Harmony made their way to ES6 (ES2015). Through browser vendor collaboration and the efforts of the working group, ES3.1 was released as ES5 in 2009 ten years after ES3, and the official ISO standard was published in 2011 after completing its fast-track process. \cite{wirfs2020javascript}

\subsection{Challenges of ES3.1}

In 2007, the main browsers in use were Internet Explorer, Mozilla Firefox, Opera, and Apple Safari. Out of these, Internet Explorer had developed a reputation for not following the ES3 specification in its implementation of JScript, and the ES3.1 working group had to address the issue. The resulting deviation document helped to understand which issues were unique to Internet Explorer and which occurred broadly and the information helped to understand where to improve the implementation of Microsoft and where to specify the standard better. \cite{wirfs2020javascript}

Given some browsers already implemented their own syntax extensions to JavaScript, the working group agreed on a rule of "3 out of 4" where the commonly implemented feature or behavior should be standardized. There was also a clear understanding that the standard should not accept changes that can break existing websites. It turns out that the rule of "3 out of 4" was enough because changing any feature with less support would not matter given the low adoption of the features. The observation makes sense given features adopted only by or two major browsers aren't likely to be used by developers as developers often prefer to support a large variety of browsers making the features niche by definition. \cite{wirfs2020javascript}

\subsection{ECMAScript conformance test suite}

Compared to other languages, ECMAScript specification had to be more strict to disallow implementation-specific variation and therefore it had to be defined better than other languages. The problem was that around the time ES5 was finalized, there was no official conformance test suite for it and as a result, one was developed by Microsoft under the name ES5conform. At the same time, Google published its own ES3 test suite, developed for its V8 JavaScript engine, known as Sputnik. These two test suites became Test262 in 2010 and the test suite became an integral part of the standardization effort from there on\footnote{As shown by JSCert \cite{bodin2014trusted, gardner2015trusted}, specifying a language through tests has its problems. JSCert is a formal, Coq-based specification of ECMAScript and during the implementation of the tool, several issues were discovered in Test262 itself.}. \cite{wirfs2020javascript}

\subsection{Main features of ES5}

As mentioned, ES5 was a continuation of the ES3.1 effort and as a result, it was a conservative release. It included the following main features \cite{dao2020nature}: strict mode, additional array and object methods, and native support for JSON. The idea of strict mode was to allow developers to use a stricter subset of the language to follow good practices and avoid common pitfalls \cite{wirfs2020javascript}. From the developer's point of view, it can be enabled by using a \textbf{"use strict;"} kind of statement that tells the interpreter to follow strict semantics for the following block or the entire file \cite{wirfs2020javascript}. The strict mode was a way to build a stricter version of JavaScript without breaking the language. The added array and object methods were standardized based on convenience and common programming patterns \cite{dao2020nature}. The added JSON (JavaScript Object Notation) support gave a data interchange format useful for definitions and especially server communication \cite{dao2020nature, wirfs2020javascript}. The new features were not earth-shattering but they improved the language. It was up to the next version, ES6, to bring bigger changes. With the intermediate ES5, standardization work was back on track after a long pause in releases.





\subsection{Discussion}

ES5 was an important release in the sense that it was able to deal with the \textbf{weight of the past} in a way that the failed ES4 was not. The failure of ES4 represented \textbf{push of the present} and it also gave \textbf{pull of the future} in the sense that it shaped the roadmap for the language and its development. During the ES5 release, the shape of the standardization effort changed due to the introduction of a conformance test suite. An official test suite made life easier for both those that specified the standard and also for those who had to implement it. The test suite became a good way to benchmark the quality of the implementations and measure their parity. Given how the web platform works, that was particularly important as it gave confidence in the interpreters and helped developers ship code on top of it.

Even though ES5 took ten years to materialize since the earlier release of the standard, it shows how important it was for TC39 to move to work in an iterative manner that respects backward compatibility over redesigning the language and addressing its flaws. Many of the features that made it to ES5 had been derived from a concrete need in the growing space of web applications and that turned out to be the way JavaScript evolved in the subsequent releases as well.

Due to dissatisfaction with JavaScript syntax, projects such as \href{https://coffeescript.org/}{CoffeeScript} (2009), began to appear \cite{maccaw2012little}. The development represents \textbf{revolution versus evolution} in \cite{shapiro1999information}'s model and as we will see in the next section, the innovation was feeding the development of the standard.


\section{ES6 and beyond}

The failure of ES4 led to a small incremental release in the form of ES5. The more ambitious features were left for ES6 (ES2015) driven by the Harmony project within TC39 \cite{wirfs2020javascript}. It is good to keep in mind that by ES6 was released in 2015, the web had evolved a lot as an industry, and web applications had started to become mainstream with the emergence of social media. Applications grew in complexity each year and it was up to the standards to respond to the increasing requirements to cover gaps discovered by developers.

\subsection{Development of ES6}

With the emergence of the Harmony project in TC39, the group adopted a new way of working in 2008 where the target was to capture new ideas either through a mailing list or through meetings. For any new idea, the author should write down the design to TC39 wiki as a strawman that can then be presented to the group and then either abandoned or improved upon. Out of these discussions, TC39 defined goals for ES6 in 2009 and set up a goal statement that gave a framework for the standards development. The group adopted a champions model where an individual member or a small group was responsible for a specific feature and it was up to them to refine the feature until it was ready to be integrated into the standard to avoid the trap of design by the committee. \cite{wirfs2020javascript}

Through the strawman and champions process, the group was able to capture over 100 strawman proposals and 17 approved proposals by May 2011. Given the scope of the proposed changes and the need for backward compatibility, the group had to define how to deal with the problem and it was decided that the design should avoid breaking changes altogether. \cite{wirfs2020javascript}

It took until 2015 to complete the standard and that is when ES6, also known as ES2015, was released. It was the largest change to the standard to date and it included features such as classes, proxies, modules, arrow functions, and many other improvements \cite{wirfs2020javascript}. The question was, how to access the features before the browsers support them and enough users have installed the browser. The solution to this problem came in the form of transpilers.

\subsection{JavaScript as a compilation target}

Transpilers are a form of compilers that accept code as an input and emit code as an output. In other words, transpilers can be used to take ES6 code and emit ES3-compatible code for example. For features that have runtime requirements, additional code is provided to make them work. On the syntax level, the code is compiled to the older syntax of the target platform. Transpilers include Mozilla's since deprecated \href{https://github.com/mozilla/narcissus}{Narcissus}, Google's \href{https://github.com/google/traceur-compiler}{Traceur} (deprecated as well), the popular community-maintained \href{https://babeljs.io/}{Babel}, and Microsoft's \href{https://www.typescriptlang.org/}{TypeScript} which provides static type checking on top of JavaScript \cite{wirfs2020javascript}.


\href{https://coffeescript.org/}{CoffeeScript} from 2009 deserves an honorary mention as it provided inspiration for JavaScript syntax introduced with ES6 \cite{maccaw2012little}. As a project, CoffeeScript addressed particularly syntax limitations of JavaScript and gave the idea of what could be. To quote \cite{maccaw2012little}, CoffeeScript was very succinct and due to this reduced code by a third to a half compared to the original pure JavaScript. With the improved standards, the usage of CoffeeScript went down. At the same time, the popularity of TypeScript went up as it addressed a major concern of developers, static typing, and helped to improve software quality as a result \cite{bierman2014understanding}\footnote{The popularity and impact of TypeScript can be seen on a standardization level as there is an ECMAScript proposal related to \href{https://github.com/tc39/proposal-type-annotations}{type annotations}.}.

Transpilers were an important set of tools for the community as they allowed developers to use new features without having to wait. Better yet, transpilers let developers experiment with potential future features and that way stimulated language design. They allowed rapid deployment of ES6 and have become a part of toolchains used by many developers.

\subsection{SmooshGate - an example of a problem working through proposals}

The current way of working through feature proposals is not without its problems as illustrated by the so-called SmooshGate in 2018 given feature proposals have the potential to break the web if approved without proper consideration \cite{chromeSmooshGateChrome}. In the case of SmooshGate, this happened in a nightly build of Mozilla Firefox as it had implemented a new, proposed \textbf{Array.prototype.flatten} behavior and at least one major website broke in the browser as a result alerting TC39 to change the naming before including the feature to the standard \cite{chromeSmooshGateChrome}.

The problem was that given JavaScript allows adding new methods to its objects, popular libraries, most notably MooTools in 2007 used this technique and that is a problem for standardization. MooTools had implemented \textbf{Array.prototype.flatten} already and in a different way than in the standard, and worse yet, in such a way that broke JavaScript extension mechanisms. To mitigate the damage to existing websites, a new name for \textbf{flatten} had to be found and one of the proposals was \textbf{smoosh}, hence the name of the event \cite{chromeSmooshGateChrome}. The final name that was chosen was simply \textbf{flat}.

\subsection{Discussion}

Although it took a while to complete, ES6 was a great success from the community's perspective. The success was further enabled by the emergence of transpilers which allowed developers to leverage the new features before the browsers supported them or before their users had updated their browsers. As a release, it was important and it included many features the standard had been missing since the beginning. Many of the features planned for ES4 made their way to ES6 and the standard has been refined further since on a yearly basis. These yearly releases are named after the year (i.e., ES2023) and let the standard improve at a steady pace while avoiding problems related to big releases that require a lot of time to publish.

By ES6, TC39 had found a good way to work and although there was \textbf{weight of the past} in the form of not being able to break functionality, it was well acknowledged. There was a clear \textbf{push of the present} in the form of multiple feature proposals by the growing community that wanted to contribute back to language. The \textbf{pull of the future} was stronger than ever as with the development of JavaScript, it was clear that it would form the backbone of modern web applications and therefore was an integral part of the modern web. It can be argued that this was not the only way it could have worked out and it is far from the original vision where JavaScript merely complemented Java. Instead, it became a whole programming language of its own capable of replacing Java even on the backend through offerings like \href{https://nodejs.org/en/}{Node.js} or \href{https://deno.land/}{Deno}.

With ES6 and versions beyond it, signs of standards wars are visible again as programming languages, such as TypeScript, challenge its hegemony. In \cite{shapiro1999information}'s model, this represents \textbf{revolution versus evolution} as newcomers retain compatibility with JavaScript but JavaScript cannot become compatible with its competitors although ideas might flow back to it through standardization efforts.



\section{Discussion} \label{sec:discussion}

As a language, JavaScript has gone beyond its initial design as the initial version was famously created in ten days during the 90s to enable client-side interactivity on websites. The development of JavaScript gave a competitive advantage to Netscape Communications it was able to leverage in the early stages of the browser wars. To catch up, Microsoft was forced to implement JScript, a compatible implementation, in its competing Internet Explorer browser. The reverse engineering effort led to pressure to standardize and as a result, a technical committee (TC39) and standardized version known as ECMAScript was created.

To address the research question, "What was the impact of standardization in the evolution of JavaScript?", Table \ref{table:eras} captures the primary developments and their conditions.

\begin{table}[ht]
 \caption{Eras of JavaScript}
 \begin{tabular}{ |p{1.9cm}|p{2.5cm}|p{2.0cm}|p{2cm}|p{2cm}|p{1.8cm}|p{1.8cm}| }
 \hline
Era & Stakeholders & Push of the present & Pull of the future & Weight of the past & Standards war & Outcome \\
 \hline
Creation \mbox{(1995-1996)} & Netscape and Microsoft & Demands of the interactive web & Vision of client-side combined with server-side code & High-level compatibility with Java & Rival evolutions & First implementation, first competitor aiming for compatibility \\
 \hline
Standardi\-zation \mbox{(1996-1997)} & Initial TC39 members & The need for a standard & Vision for an interoperable language for the web & Behavior of the existing implementations & Ended due to alignment & Creation of ECMAScript standard and TC39 \\
 \hline
From ES1 to ES3 \mbox{(1997-1999)} & TC39 members & Integration of browser-specific extensions & Shared vision of scripting on the web & Backward compatibility & None & Improved standard \\
 \hline
ES4 \mbox{(1999-2008)} & TC39 members & Fixing the language & Visions of scripting on the web & Backward compatibility & Rival evolutions & Failure, reformation of TC39, Harmony \\
 \hline
ES5 \mbox{(2008-2015)} & TC39 members & Failure of ES4 & Roadmap shaped by the failure of ES4 & Backward compatibility & Revolution versus evolution & Successful release of ES5, improved way of working for TC39 \\
 \hline
ES6 and beyond (2015-) & TC39 members and web community & Feature proposals from the community & Clear vision of JavaScript as a part of the modern web & Backward compatibility & Revolution versus evolution & Champions model, yearly snapshots \\
 \hline
 \end{tabular}
 \label{table:eras}
\end{table}

The impact of standardization on JavaScript can be seen on many levels. In retrospect, not everything went as well as it could have as there were false starts, as in the form of ES4, and it took a while for TC39 to find good ways of working. At the same time, the creation of the ECMAScript standard gave a good way for browser vendors to collaborate and it can be argued that without this collaboration JavaScript would not have its current status. Later on, with ES6, it became easier for the broader community to participate in the effort and yearly snapshots of the standard to keep it relevant.

In terms of standards wars, JavaScript followed a path where first there was rivalry, then a period of peace before rivalry, and finally revolution versus evolution where challengers treat JavaScript as a target language while providing features beyond it to entice users to their offering. TypeScript is a good example of such a language as it addresses typing, a main pain point of JavaScript, and has therefore gained a significant amount of users.

Another promising direction is a \href{https://www.w3.org/}{W3C} standard called \href{https://webassembly.org/}{WebAssembly} which sidesteps JavaScript for computation intensive tasks \cite{haas2017bringing}. WebAssembly provides a compilation target for many existing programming languages and therefore increases the attractiveness of the web platform as more code can be run on top of it.

\subsection{Alternative future led by Microsoft}

The future could have been different if Microsoft had captured the browser market before Netscape as then it could have leveraged its existing technologies, such as Visual Basic, on top of the web platform. There were attempts in this direction with Visual Basic Script and ActiveX but they failed. As Netscape was first, it was able to force Microsoft to follow its lead. If the roles were reversed, likely JavaScript would not exist and we might use some technology pioneered by Microsoft instead.

In the early phase of development, standardization was a forced move to allow collaboration and competition at the same time. It can be argued that had Microsoft gained an early lead in the web space, we might have a standard based on their technology or alternatively, we could have headed to a configuration where proprietary technologies prosper and the open web would not exist as we know it.

\subsection{The costs and payoffs of standardization}

Although agreeing on a standard fixed the playing field to some extent, it also left space for innovation outside of the official specification in the form of extensions and intangible features, such as performance as shown later by Google's Chrome browser and its V8 JavaScript engine. Later on, V8 came to form the core of Node.js and other key technologies that enable the usage of JavaScript on the backend. Even many desktop applications are developed using JavaScript these days and it has become a truly ubiquitous language.

If Netscape had kept JavaScript as a proprietary technology and forced Microsoft to keep reverse engineering it, that would not have been good for the web platform as a whole and likely it would have hindered its growth as fewer players could access the market due to the higher barrier of entry. At the same time, assuming Microsoft gained more market share, as they did in the end, Microsoft could have done the same and pushed their proprietary solution or gone the way of standardization assuming the benefits were considered higher than the drawbacks.

\subsection{Standards allowed to capture browser extensions}

Given different browser vendors were implementing JavaScript extensions of their own, early standardization efforts gave a way to capture the most useful developments to the ECMAScript standard as shown by the progress made with ES3. On top of this, standardization gave a way to specify earlier undefined behavior and harmonize browser behavior. The benefit of this was a stronger web platform that was better for developers as the challenge with JavaScript is that the code is interpreted in the browsers directly instead of using a prepackaged binary form that would have alleviated the issue or at least pushed it elsewhere.

\subsection{Design by committee slowed standardization}

As shown by the example of the failed ES4 standardization effort, driving the full design effort of a programming language can be challenging. Although the target of fixing the mistakes made in the design of the original language was a noble one, it was too much for TC39 to handle in this case. The important lesson for TC39 was to maintain backward compatibility to avoid breaking the web and be more mindful of its changes. Big changes were still possible but their impact had to be evaluated more carefully.

\subsection{Iterative improvements moved standardization back on track}

As shown by ES5, ES6, and releases beyond, the right way to standardize ECMAScript was an incremental one. In the current model of yearly releases, features make it to the standard each year in a steady cycle. TC39 went with a champion model that has proven to work as it allows individuals to focus on specific features while TC39 can evaluate the proposals in a larger scope and consider the interactions of the proposals on a larger scale.

The model allows developments occurring outside of ECMAScript to make their way to the language. Occasionally trends triggered by technologies such as CoffeeScript or TypeScript inspire proposals. They also have a more indirect way of influencing standards through thought leadership as programming trends influence the decision makers of TC39 as well and may make them more amenable towards proposed changes. The tendency of the developers to move towards static typing and stronger typed systems is a good recent example of this as the shift is putting pressure on the standard to shift the baseline.



\section{Conclusion} \label{sec:conclusion}

ECMAScript is a good example of a programming language where standardization helped to leverage its position. It can be argued that without early standardization, the language might not exist as we know it and we might use some other programming language on the web platform. Even with its mistakes and struggles during standardization, JavaScript has proven to be good enough for web applications and it has gone far beyond its initial design envelope where it was supposed to complement Java, a "real" programming language.

TypeScript, a JavaScript dialect by Microsoft, provides a counterpoint to standardization. Although TypeScript is increasingly popular as it provides strong benefits for its users, there is no standard for the language. The lack of standardization allows the language to move fast and it is not uncommon for a new release to break existing behavior in favor of improved syntax or stricter type-checking for example. It can be argued that at least for now, it is good that TypeScript does not have a standard as standardization would affect its velocity as a technology. At the same time, the lack of standards makes it more complex for third parties to create tools around TypeScript but until now this has not been a major issue.

There are early signs that TypeScript may have an impact on ECMAScript as proposals related to typing are considered. Although ECMAScript may become more strongly typed, it feels unlikely that TypeScript would disappear entirely but at the same time, ECMAScript-level changes would give a better compilation target for it as then more type information could be retained during the transpilation process. That in turn could lead to other benefits, such as stronger type guarantees during execution and potentially better performance due to fewer amount of type inferences required.

TypeScript and its lack of standardization would be a good research topic itself. A larger question is related to programming languages in general as not all popular languages, such as Perl or PHP\footnote{Instead of an official standard, PHP has a de facto standard and does not have a standardization organization behind it.}, have an official standard. The question is, what benefits and drawbacks does a non-standardized model provide in contrast to a model driven by a standard as seen with ECMAScript and its popularity?

On a broader level, the case of ECMAScript shows the power of standardization as it enabled competitors to collaborate and build a large market for all. ECMAScript has become an integral part of the modern web and it is one of the key technologies enabling web applications. Similar benefits may be possible to attain in upcoming themes related to aspects such as data, artificial intelligence, or machine learning. As standards are created, they open new, and sometimes unexpected, ways of collaboration while pushing the baseline of the technology higher allowing competition and innovation outside of the standardized portions of technology that in turn makes its way back to the standards over time.

\bibliographystyle{apalike}
\bibliography{references}

\end{document}